\documentclass[9pt,twocolumn,twoside]{osajnl}

\journal{ao} 

\setboolean{shortarticle}{false} 
\usepackage{csquotes}

\title{Static spectropolarimeter concept adapted to space conditions and wide spectrum constraints}

\author[1,2,*]{Martin Pertenais}
\author[1]{Coralie Neiner}
\author[1]{Pernelle Bernardi}
\author[1]{Jean-Michel Reess}
\author[2,3]{Pascal petit}

\affil[1]{LESIA, Observatoire de Paris, PSL Research university, CNRS, Sorbonne Universit\'es, UPMC Univ. Paris 06, Univ. de Paris-Diderot, Sorbonne Paris Cit\'e, 5 place Jules Janssen, 92190 Meudon, France}
\affil[2]{Universit\'e de Toulouse; UPS-OMP; IRAP Toulouse, France}
\affil[3]{CNRS; IRAP; 14 avenue Edouard Belin, 31400 Toulouse, France}

\affil[*]{Corresponding author: martin.pertenais@irap.omp.eu}

\dates{Compiled \today}

\ociscodes{(120.5410) Polarimetry, (260.5430) Polarization, (260.1440) Birefringence, (120.6085) Space instrumentation, (350.1260) Astronomical optics, (350.1270) Astronomy and astrophysics.}

\doi{\url{http://dx.doi.org/10.1364/ao.XX.XXXXXX}}

\begin{abstract}
The issues related to moving elements in space and instruments working in broader wavelength ranges lead to a need for robust polarimeters, efficient on a
wide spectral domain, and adapted to space conditions. As part of the UVMag consortium, created to develop spectropolarimetric UV facilities in space, such as the Arago mission project, we present an innovative concept of static spectropolarimetry. We studied a static and polychromatic
method for spectropolarimetry, applicable to stellar physics.  Instead of modulating the polarization information temporally, as usually done in spectropolarimeters, the modulation is performed in a spatial direction, orthogonal to the spectral one. Thanks to the proportionality between phase retardance imposed by a birefringent material and its thickness, birefringent wedges can be used to create this spatial modulation. The light is then spectrally cross-dispersed, and a full-Stokes determination of the polarization over the whole spectrum can be obtained with a single-shot measurement. The use of Magnesium Fluoride wedges, for example, could lead to a compact, static polarimeter working at wavelengths from 0.115~$\mu$m up to 7~$\mu$m. We present the theory and simulations of this concept, as well as laboratory validation and a practical application to Arago.
\end{abstract}

\setboolean{displaycopyright}{true}

\begin{document}

\maketitle
\thispagestyle{fancy}
\ifthenelse{\boolean{shortarticle}}{\abscontent}{}


\section{Introduction}
\label{intro}
\subsection{Context}
Analysing the polarization state of the light provides crucial information in many different areas, for example in remote sensing for medical or biophysics issues, military applications, atmosphere studies, and of course astronomy \cite{Workshop}. The light that we receive from objects in our universe can be polarized through several physical processes. Polarized light can be observed from, e.g., the interstellar medium in our galaxy, other galaxies, quasars, or supernovae. In stellar physics, the Zeeman effect allows us to determine the polarization state of the light over the spectrum, which enables the measurement of the magnetic field of stars and provides constraints on circumstellar environments. For observations of the Sun, both the Zeeman and Hanle effects are used to determine the local magnetic field in and around the Sun. 

In this work, we aim to measure and characterise magnetic fields of stars and polarization from the stellar surroundings, e.g., the disk or stellar wind. However, it is not yet possible to resolve the surface and surroundings of stars in the same way as is done for the Sun, with the stellar target apppearing only as a point source in observations. Nevertheless, analysing the polarization of the light emitted by the star in its photospheric absorption spectral lines offers an easy way to measure the strength of the longitudinal stellar magnetic field, as it is directly proportional to the intensity of the polarized light. The presence of a magnetic field manifests as characteristic Zeeman signatures in
the polarized part of the light, specifically in the spectral photospheric lines.
Measuring this magnetic signature as the star rotates allows us to
characterise the full magnetic field at the surface of the star and, in
particular, its polar strength, obliquity with respect to the rotation axis, and
surface configuration. Similarly, polarization in wind-sensitive UV
resonance lines, or in emission lines originating from disks or
choromospheres, allows us to study the circumstellar environment.
While some science cases may be tackled by measuring circular (Stokes V)
polarization only, others require linear (Stokes Q and U) polarization. However, it
has been shown that 3D magnetic mapping of stars and their environment is more
precise when all Stokes parameters are used simultaneously
\citep[e.g.,][]{KochukhovWade2010, Rosen2015}. In addition, to allow a space
mission to address various science objectives, it is desirable to provide
measurements of both circular and linear polarization.\\

To characterize and quantify the polarization of light, we need to measure its polarization state all over the entire spectrum, and with a high spectral resolution. However, as common detectors are only sensitive to the intensity of the light and not to its polarization, we have to build polarimeters to encode the polarization state information into the intensity of the light.  The next section describes the method to determine the polarization states of the light.

\subsection{Formalisms and tools}
\label{tools}

\subsubsection{Stokes and Mueller formalisms}
To go into more details, the Stokes and Mueller formalisms are briefly introduced in this section. They are well documented in the literature, for example in the books "Polarized light" from W. A. Shurcliff \cite{Shurcliff} or "Birefringent Thin Films and Polarizing Elements" from I. J. Hodgkinson \cite{Hodgkinson}. 

Considering a monochromatic electromagnetic wave propagating in the $\overrightarrow{z}$ direction, the transverse components of its electric field $\overrightarrow{E}$ are defined by:
\begin{equation}
\left \{
\begin{array}{c}
	E_x =\xi_{x}.\cos(\omega t-kz)  \\
	E_y =\xi_{y}.\sin(\omega t-kz+\varphi)
\end{array}
\right .
\end{equation}

with $\omega$ its angular frequency, $t$ the time, and $k$ the wave number related to its wavelength.

The Stokes vector, giving complete information of the light (intensity and polarization state), is then defined using the amplitudes $\xi_x$ and $\xi_y$ of the electric field and the phase shift $\varphi$ between its two components:
\begin{equation}
\left (
\begin{array}{c}
	I \\
	Q \\
	U \\
	V \\
\end{array}
\right )
=
\left (
\begin{array}{c}
	\xi_{x}^2+\xi_{y}^2 \\
	\xi_{x}^2-\xi_{y}^2 \\
	2\xi_{x}\xi_{y}\cos \varphi \\
	2\xi_{x}\xi_{y}\sin \varphi \\
\end{array}
\right )
\end{equation}

The first Stokes parameter I defines the total intensity of the light, while the intensity of the polarized part of the light is defined by $ \sqrt{Q^2+U^2+V^2} $. The ratio of these two values defines the degree of polarization of the light $p$:
\begin{equation}
p=\frac{\sqrt{Q^2+U^2+V^2}}{I}
\end{equation}

The Mueller (4x4) matrix is  introduced to link the input and output Stokes vectors after an optical element. Adding several optical elements in the system corresponds to multiplying the Mueller matrices of the elements. Therefore, the input Stokes vector we want to determine is linked to the output Stokes vector we measure with our polarimeter through the total Mueller matrix of the system $M_{tot}$:
\begin{equation}
\overrightarrow{S_{out}}=M_{tot} \cdot \overrightarrow{S_{in}}
\end{equation}

Developing a polarimeter consists in finding the optimal combination of optical elements to determine the input Stokes parameters from the measured output ones. As detectors are only sensitive to the intensity, the goal is that the Mueller matrix transfers the full input Stokes vector into the first Stokes parameter I of the output Stokes vector. Mathematically speaking, we aim at a final equation of the type $I_{out}=x_I.I_{in}+x_Q.Q_{in}+x_U.U_{in}+x_V.V_{in}$, with $x_I, x_Q, x_U$ and $x_V$ combinations of some elements of the total Mueller matrix, depending on the characteristics of the optical elements. As a consequence, a modulation of the polarization states is needed, because we have a single equation with four unknowns, I, Q, U, and V. 

\subsubsection{Modulation of the polarization state}
A typical solution is to modulate the polarization state temporally, by rotating some optical elements in the system to solve the above equation. However, in space missions, rotating or more generally moving parts should  be avoided if possible in the instrument, as they are automatically considered as single point failure (i.e., risk items for the mission). Using liquid crystal switching systems could be a solution to modulate temporally the polarization state without rotating optical parts \cite{liquidcrystal}. Unfortunately, liquid crystals are not efficient in wide spectral ranges and cannot be used in the far-ultraviolet (FUV) region, below 250 nm. In addition, temporal modulation provides successive rather than simultaneous measurements of the Stokes Q, U and V parameters. Instead of temporal modulation, polarization can be modulated spectrally \cite{spex} or spatially using channeled polarimetry or division of amplitude. 

Our goal is to create an efficient static spectropolarimeter over a wide spectral range, which provides the full Stokes vector simultaneously. This means we need to modulate the polarization spatially, over the wavelength range. As we use a spatial dimension for the modulation, this method can only be applied to point sources, such as stars. This solution imposes a variable retardance to the beam in one spatial direction to obtain several polarization states at the same time, without moving any part of the system. Ideally, the spatial direction in which the polarization will be modulated must be orthogonal to the spectral direction, so that we obtain a 2D array containing all the polarization and spectral information of the entrance light. 

Birefringent wedges, generally known as polarization scramblers, can be used to create this spatial modulation \cite{Sparks}. Combined with a linear polarizer and a spectrograph, we can create a static spectropolarimeter working over a wide spectral range. The only limitations of this system are the efficiencies of the materials and the design of the spectrograph. 

\subsubsection{Material}

\begin{figure}[!t]
\centering
\includegraphics[width=\linewidth]{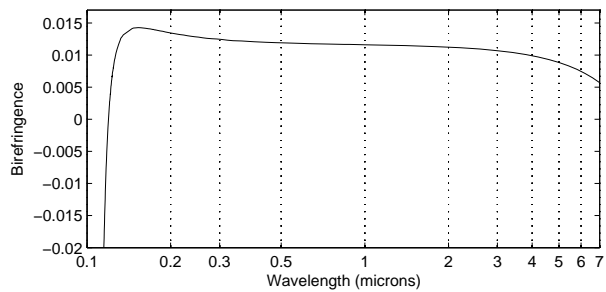}
\caption{Birefringence curve of MgF$_2$ from FUV (115~nm) to IR (7~$\mu$m) \protect\cite{Chandra68, Chandra69, Dodge, clasp}.}
\label{birefringence}
\end{figure}

Materials such as quartz, sapphire, or calcite are good candidates to create a polychromatic spectropolarimeter, because of their transparency and birefringent characteristics. However, for an instrument covering a wavelength range from FUV (0.115~$\mu$m) to infrared (IR) (7~$\mu$m), the only available transparent and birefringent material is Magnesium Fluoride (MgF$_2$). In this work, we thus consider MgF$_2$ wedges.

The birefringence curve of MgF$_2$ over the wavelength range 0.115 to 7~$\mu$m is displayed in Fig.~\ref{birefringence}.

The measurements for this plot were obtained by Chandrasekharan in 1968 and 1969 \cite{Chandra68, Chandra69} and Dodge in 1984 \cite{Dodge}. For the space mission CLASP, aiming at solar spectropolarimetry at Ly$\alpha$, a short part of the birefringence curve has been measured again in 2013 \cite{clasp}. The various data were combined to obtain the complete plot of the birefringence of MgF$_2$ over the [0.115:7]~$\mu$m spectrum (Fig.~\ref{birefringence}).

The birefringence of MgF$_2$, $\Delta$n, defined by the difference between the extraordinary refractive index $n_e$ and the ordinary one $n_o$, only varies between 0.005 and 0.015 from 122~nm to 7~$\mu$m. It is transparent down to 115~nm, but its birefringence drops off dramatically below 140~nm, goes to zero around 119.5~nm, and inverts its sign below this wavelength. The consequence is that it is very difficult to use MgF$_2$ to measure polarization states between 119 and 120~nm, as the value of the birefringence is too low. Nevertheless,  MgF$_2$ offers the best opportunity to create a polychromatic spectropolarimeter including the FUV domain.\\

In the next section of this paper (Sect.~\ref{theory}), we describe the conceptual theory, calculations, and simulations of a spectropolarimeter using spatial modulation with MgF$_2$ wedges. After applying the method for the design of an instrument for the Arago space mission project (Sect.~\ref{arago}), the laboratory implementation section (Sect.~\ref{manip}) shows the practical results and comparison with simulations.


\section{Theory and simulations}
\label{theory}

\subsection{Global principle} \label{global}
To achieve a spatial modulation of the polarization, we need to use a component that gives a variable phase retardance between the extraordinary and the ordinary part of the light. As shown in Eq.~\ref{retardance}, the retardance of a plate is directly proportional to its thickness. If the thickness $d$ of the plate varies continuously in a certain direction, the retardance $\Phi$ also varies accordingly.

\begin{equation}
\label{retardance}
\Phi=\frac{2\pi\Delta n}{\lambda} \cdot d
\end{equation}

We use a birefringent wedge (polarization scrambler) of apex angle $\xi$ and birefringence $\Delta n =n_e-n_o$. The wedge is illuminated by a collimated beam coming from a point source. The link between the retardance and the height $x$ along the vertical direction of the wedge is given in Eq.~\ref{retardance2}, with $x=0$ at the top of the wedge:

\begin{equation}
\label{retardance2}
\Phi(x,\lambda)=\frac{2\pi\Delta n(\lambda)\tan{\xi}}{\lambda} \cdot x
\end{equation}

To ultimately create a spectropolarimeter, we then disperse the light using, for example, a grating, which causes dependence of the retardance on both the height $x$ and the wavelength $\lambda$.

\begin{figure}[tbp]
\centering
\includegraphics[width=\linewidth]{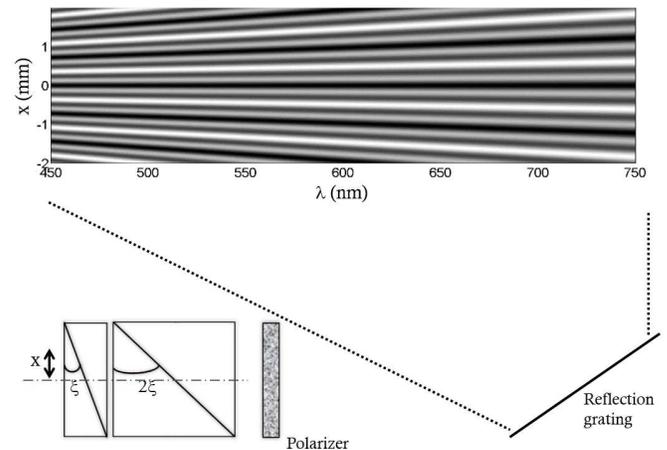}
\caption{Global principle of the measurement. The light travels through two double-wedge birefringent components, a linear polarizer, and finally a grating to create the spectrum. A 2D image is obtained with the spectral information in the horizontal direction and the polarization modulation in the vertical one.}
\label{principe}
\end{figure}

Combining a single birefringent wedge with a linear polarizer cannot measure the full Stokes vector, but only two of the three Stokes parameters in addition to the intensity I (see more details in Sparks et al. (2012) \cite{Sparks}). To solve this problem and obtain a full Stokes determination, we need to add a second birefringent prism with a slope (or apex angle) twice as large as the one of the first wedge. In order to make a rectangular component, which is easier to handle, and to introduce a symmetry around the optical axis, the two wedges can be doubled with the same material, but with an orthogonal fast axis angle. In this case, on the optical axis, there is no birefringent effect and the retardance is zero. This configuration is shown in the bottow left part of  Fig.~\ref{principe}, and in more detail in Fig.~\ref{wedge}. 

As detailed in the next subsection, this configuration of wedges followed by a linear polarization analyzer enables us to encode the polarization information into the output intensity. The light is then dispersed with a grating and focused onto a detector. The resulting image contains all the needed information: a classical spectrum with, for each wavelength, the full Stokes information encoded in the orthogonal direction of the spectrum (see Fig.~\ref{principe}).

It is possible to use a dual-beam analyzer instead of the linear polarizer. In this case, the signal-to-noise ratio (SNR) increases by a factor of $\sqrt{2}$, and some systematic errors (due to illumination inhomogeneity for instance) can be suppressed. Unfortunately, such a system would double the size of the required detectors, which is a critical issue as shown in Sect.~\ref{arago}. 

The SNR obtained with this concept can be compared to other techniques and, in particular, to a more usual polychromatric rotating modulator. For this
comparison, we consider the dual-beam option, as it is a standard technique for spectropolarimeters with rotating plates. The polychromatric rotating modulator we consider is composed of 3 quasi-zero order MgF$_2$ plates at 6 different angular positions, optimized to maximize the extraction efficiencies over the spectral range \cite{pertenais}. For the static spectropolarimeter presented here, the flux at each wavelength is spread along a column of pixels,
while for the rotating system it is divided in several sub-exposures. The ratio of the SNR resulting from the two spectropolarimeters is equal to the square
root of the ratio between the number of pixels used for the concept presented in this article and the number of sub-exposures for the rotating technique.


\subsection{Mueller matrix calculation}
\label{mueller}

Using Mueller matrices applied to this system, this section presents in more detail how the polarization is modulated and encoded.

\begin{figure}[tbp]
\centering
\includegraphics[width=\linewidth]{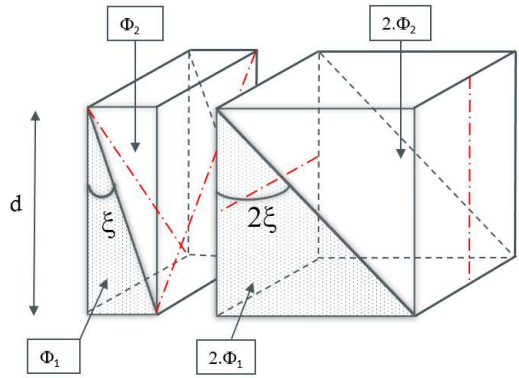}
\caption{Configuration of the birefringent wedges to obtain full Stokes measurement. The first bloc of 2 wedges has a total retardance $\phi=\phi_1-\phi_2$ and the second bloc has twice this retardance : $2\phi$. The fast axis angle values (with respect to the horizontal) for the four wedges are [$\frac{3\pi}{4}$; $\frac{\pi}{4}$; 0; $\frac{\pi}{2}$] from the left to the right (shown with red dashed-dotted lines).}
\label{wedge}
\end{figure}

Fig.~\ref{wedge} defines precisely the configuration of the birefringent wedges. The fast axis angle $\alpha_1$ of the first wedge of retardance $\Phi_1$ is set to $\alpha_1=\frac{3\pi}{4}$, and its apex angle is $\xi$. The fast axis angle of the second wedge of retardance $\Phi_2$ is set to $\frac{\pi}{4}$. The total phase retardance of the first rectangular component is then  $\Phi=\Phi_1-\Phi_2$. Because the two wedges have orthogonal fast axes,  the retardance cancels for equal thicknesses, and the global retardance on the optical axis is thus $\Phi=0$.

The second rectangular component is built in the same way, but with the fast axis angles of $0^{\circ}$ and $\frac{\pi}{2}$ and wedges with apex angle $2\xi$.

The retardance $\Phi$ is directly calculated using Eq.~\ref{retardance2}:

\begin{align}
\Phi(x,\lambda)&=2\pi\tan{\xi} \cdot \frac{\Delta n(\lambda)}{\lambda} \cdot x - 2\pi\tan{\xi} \cdot \frac{\Delta n(\lambda)}{\lambda} \cdot (d-x) \nonumber \\
&=2\pi\tan{\xi} \cdot \frac{\Delta n(\lambda)}{\lambda} \cdot (2x-d)
\end{align}

The above considerations are made with $x=0$ at the top of the wedges. Changing the variable to have $x=0$ on the optical axis (as shown in Fig.~\ref{principe}) leads to a global retardance of:

\begin{equation}
\label{phi}
\Phi(x,\lambda)=4\pi\tan{\xi} \cdot \frac{\Delta n(\lambda)}{\lambda} \cdot x
\end{equation}

The Mueller matrix of the first rectangular component $D_1$ is the product of the Mueller matrices of the first wedge $D_{\frac{3\pi}{4}}$ and the second wedge $D_{\frac{\pi}{4}}$ : $D_1=D_{\frac{\pi}{4}} \cdot D_{\frac{3\pi}{4}}$.

\begin{align}
D_1&=
\left(
	\begin{matrix}
	1 & 0 & 0 &0 \\
	0 & \cos{\Phi_2} & 0 & -\sin{\Phi_2} \\
	0 & 0 & 1 & 0 \\
	0 & \sin{\Phi_2} & 0 & \cos{\Phi_2} \\
	\end{matrix} 
\right)
\cdot
\left(
	\begin{matrix}
	1 & 0 & 0 &0 \\
	0 & \cos{\Phi_1} & 0 & \sin{\Phi_1} \\
	0 & 0 & 1 & 0 \\
	0 & -\sin{\Phi_1} & 0 & \cos{\Phi_1} \\
	\end{matrix} 
\right) \nonumber \\
&= \left(
	\begin{matrix}
	1 & 0 & 0 &0 \\
	0 & \cos{\Phi} & 0 & \sin{\Phi} \\
	0 & 0 & 1 & 0 \\
	0 & -\sin{\Phi} & 0 & \cos{\Phi} \\
	\end{matrix} 
\right)
\end{align}

For the second component, the calculus is the same with $D_2=D_{\frac{\pi}{2}} \cdot D_{0}$. The result is directly shown here:
\begin{equation}
D_2= \left(
	\begin{matrix}
	1 & 0 & 0 & 0 \\
	0 & 1 & 0 & 0 \\
	0 & 0 & \cos{2\Phi} & \sin{2\Phi} \\
	0 & 0 & -\sin{2\Phi} & \cos{2\Phi} \\
	\end{matrix} 
\right)
\end{equation}

The global Mueller matrix $D$ of the polarimeter modulator is developed in Eq.~\ref{Mueller}:
\begin{align}
D&= D_2 \cdot D_1 \nonumber  \\
&=
\label{Mueller}
\left(
	\begin{matrix}
	1 & 0 & 0 & 0 \\
	0 & \cos{\Phi} & 0 & \sin{\Phi} \\
	0 & -\sin{\Phi}\sin{2\Phi} & \cos{2\Phi} & \cos{2\Phi}\sin{2\Phi} \\
	0 & -\cos{2\Phi}\sin{\Phi} & -\sin{2\Phi} & \cos{\Phi}\cos{2\Phi} \\
	\end{matrix} 
\right)
\end{align}

The final step to determine the equation of the resulting intensity $I_{out}$ after passing through the polarimeter is to compute the total Mueller matrix of the polarimeter and to mutliply it by the entrance Stokes vector $( \begin{matrix}  I  & Q &  U &  V \end{matrix} )^T$. The matrix $D$ has to be multiplied by the first line of the Mueller matrix of the linear polarizer $0.5~\cdot~( \begin{matrix}  1  & \cos{2\theta} &  \sin{2\theta} &  0 \end{matrix} )$, where $\theta$ is the angle of the polarizer with respect to the horizontal. The 2D output intensity, depending on the height in the wedge $x$, the wavelength $\lambda$, and the input Stokes parameters I, Q, U and V, is described with Eq.~\ref{I}:

\begin{multline}
\label{I}
I_{out}(x,\lambda)=0.5 \cdot [I+Q \cdot \left( \cos{\Phi}\cos{2\theta}-\sin{\Phi}\sin{2\Phi}\sin{2\theta} \right) + \\ 
U\cos{2\Phi}\sin{2\theta}+V \cdot \left( \cos{\Phi}\sin{2\Phi}\sin{2\theta}+\sin{\Phi}\cos{2\theta} \right)]
\end{multline}

This equation, in combination with Eq.~\ref{phi}, provides the starting point for simulations and to the design of our laboratory implementation.

\subsection{Simulations and inversion}

For the simulations, MgF$_2$ birefringent wedges, with an apex angle $\xi=1.5^{\circ}$ are used, as well as a fixed linear polarizer with angle $ \theta=64^{\circ}$. The value of this angle was chosen so that the 3 basic 2D patterns corresponding to 100\% Q, 100\% U and 100\% V entrance polarization are different. This is not mandatory, but makes the demodulation process easier.

\begin{figure}[t]
\centering
\includegraphics[width=\linewidth]{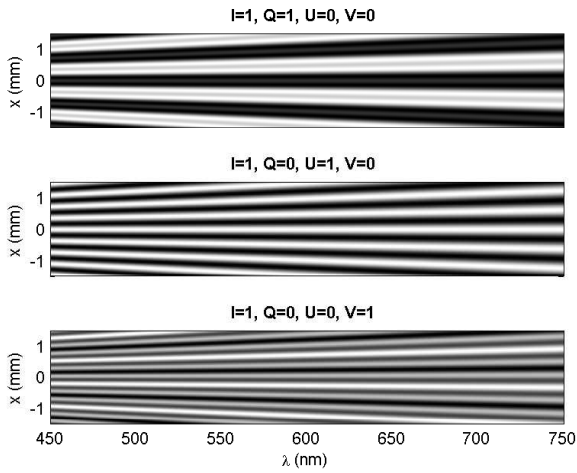}
\caption{Simulation of 2D images obtained by solving Eqs.~\ref{phi} and \ref{I} for given polarization states. The values of the constants used in this simulation are an apex angle of $\xi=1.5^{\circ}$ and a linear polarizer angle of $ \theta=64^{\circ}$. }
\label{simu}
\end{figure}

 A given entrance polarization state of the light is set: I=1 and Q=1, U=0, V=0 in the first case, Q=0, U=1, V=0 in the second case, and finally Q=0, U=0, V=1.  The value of the birefringence plotted in Fig.~\ref{birefringence} is used to calculate the phase retardance at each wavelength $\lambda$ and height $x$, using Eq.~\ref{phi}. Finally, Eq.~\ref{I} is solved for a spectral range from 450 to 750~nm with wedges of 3-mm height ($x$ varying from -1.5 to 1.5~mm). The 2D output intensities, with respect to the wavelength and the height $x$, are shown in Fig.~\ref{simu}. Three different 2D patterns are indeed obtained for the 3 cases, corresponding to 100\% Q, 100\% U, and 100\% V polarization.

These images are compared with real data obtained in the laboratory in Sect.~\ref{manip} .

In practice, when real stellar observations are obtained, Eq.~\ref{I} is used the other way around. We collect an image that corresponds to a given output intensity, and need to determine the entrance polarization state (I ~~Q~~ U~~ V) producing the image. This inversion can be performed  through a classical least-squares solving method. The parameters of the systems ($\xi$, $\theta$) and the variables ($x$, $\lambda$) have to be well known in order to achieve a precise measurement of I, Q, U, and V over the spectrum. 

\begin{figure}[t]
\centering
\includegraphics[width=\linewidth]{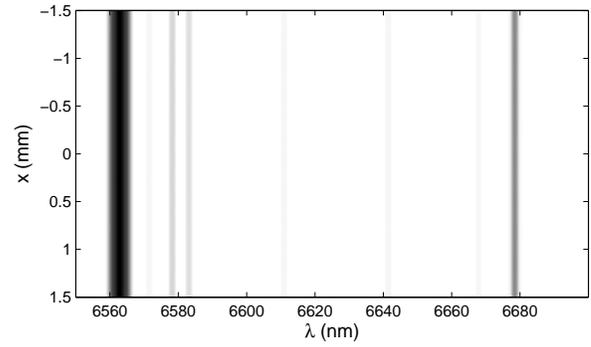}
\caption{Simulation of a 2D image obtained for a synthetic star, between 655 and 670~nm. Photospheric lines are observed, but the polarization modulation is not seen by eye.}
\label{I_simuspectre}
\end{figure}

Fig.~\ref{I_simuspectre} shows the simulation of the image we would obtain with the polarimeter described above and a high-resolution (R=65000) spectrometer, for a star with an effective temperature of $T_{\rm eff}$=30000 K, gravity of $\log g$=4.0, projected rotational velocity $v \sin i$ of 50 km~s$^{-1}$, solar abundances, a dipolar magnetic field with a strength of $B_{\rm pol}$=1000 G inclined by $\beta$=90$^\circ$ from the rotation axis. The star is seen equator-on ($i$=90$^\circ$) with the magnetic pole facing the observer (magnetic phase is zero).
The black lines seen in Fig.~\ref{I_simuspectre} are absorption lines due to the photosphere of the star. As the QUV Stokes parameters of the stars are very small compared to its intensity, no polarization modulation can be seen by eye. However, the information is hidden in the intensity, following Eqs.~\ref{phi} and \ref{I}, which can be summarized by:
\begin{multline}
\label{I2}
I_{out}(x,\lambda)=\alpha_I(x,\lambda).I(\lambda)+\alpha_Q(x,\lambda).Q(\lambda)+ \\
\alpha_U(x,\lambda).U(\lambda) +\alpha_V(x,\lambda).V(\lambda)
\end{multline}

Taking the image of Fig.~\ref{I_simuspectre} as the observed data (corresponding to $I_{out}(x,\lambda)$ in Eq.~\ref{I2}), we retrieve the Stokes parameters $I(\lambda)$, $Q(\lambda)$, $U(\lambda)$, and $V(\lambda)$ of the star over the spectrum using the least-squares method. We solve, for each wavelength, the overdetermined linear system: $I_{out}=A.S$ with $A=( \alpha_I ~~ \alpha_Q~~ \alpha_U ~~ \alpha_V ) $ and $S$ the Stokes vector we want to determine $\left( \begin{matrix}  I(\lambda)  & Q(\lambda) &  U(\lambda) &  V(\lambda) \end{matrix} \right)^T$.

\begin{figure}[!htbp]
\centering
\includegraphics[width=\linewidth]{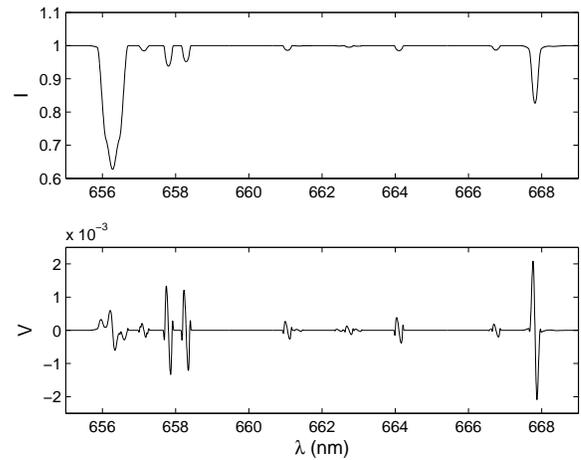}
\caption{Results of the inversion code for the image presented in Fig.~\ref{I_simuspectre}. Stokes I (top panel) clearly shows the photospheric absorption lines and Stokes V (bottom panel) shows Zeeman signatures at the corresponding line wavelengths, proof that the star hosts a magnetic field.}
\label{V_simuspectre}
\end{figure}

The results for Stokes I and Stokes V are displayed in Fig.~\ref{V_simuspectre}.

The photospheric absorption lines are retrieved in the Stokes I spectrum. The star chosen for the simulation is magnetic, which means that we should find Zeeman signatures at the position of the absoption lines in the Stokes V spectrum. Zeeman signatures are a direct proof of the presence of a magnetic field: the spectral lines split in wavelength in two components, which are left and right circularly polarized. Stokes V shows the difference between these two components, and the splitting width is proportional to the magnetic field strength. Even if the strength of the Stokes V signal is 1000 times lower than the one of I, we retrieve the Zeeman signatures in Stokes V based on the image simulation of Fig.~\ref{I_simuspectre}.

\begin{figure}[tbp]
\centering
\includegraphics[width=\linewidth]{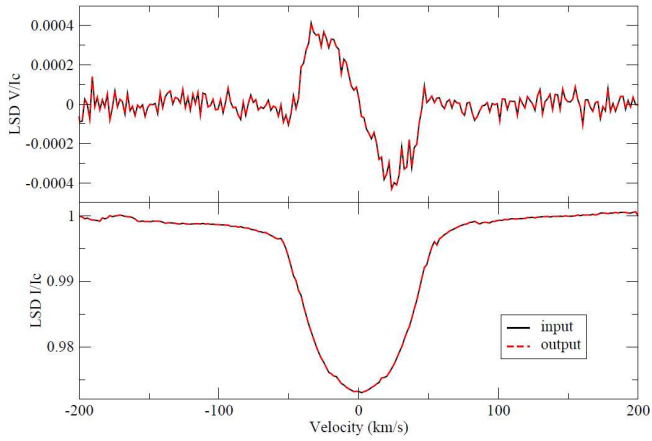}
\caption{LSD profiles calculated for Stokes I and V for the synthetic spectrum with SNR=1000 (input, solid black line) and for the I and V spectra determined with the wedges concept and the demodulation code (output, dashed red line). They are strictly identical.}
\label{LSD}
\end{figure}

To extend the analysis further, the same simulation is performed but with noise in the initial synthetic spectrum of the star. A SNR of 1000 is used with the same star as in the previous simulation. To be able to observe the Zeeman signatures, which have a similar amplitude as the noise, the Least-Squares Deconvolution (LSD) technique is applied \cite{LSD_paper}. This technique uses a Dirac comb corresponding to all the stellar spectral lines and average the Stokes I and V signals in those lines obtaining a mean pattern of the line profile and Zeeman signature. We apply this technique for all lines from 400 to 900~nm, both directly to the synthetic spectrum and to the spectrum extracted from the 2D image. The results are plotted in Fig.~\ref{LSD}. We retrieve a mean line profile and the mean Zeeman signal in Stokes V corresponding to the input magnetic field. Moreover, the pattern of the LSD profiles for Stokes I and Stokes V are exactly identical for the input synthetic spectrum and for the one determined by the demodulation code from the 2D simulation. This result proves the validity of the theoretical concept and of the demodulation code.

\subsection{Single component solution}

In some applications, in particular in astronomy, the amount of collected flux can be a crucial issue. A solution to decrease the number of photons lost through the polarimeter module is presented here.

\begin{figure}[t]
\centering
\includegraphics[width=\linewidth]{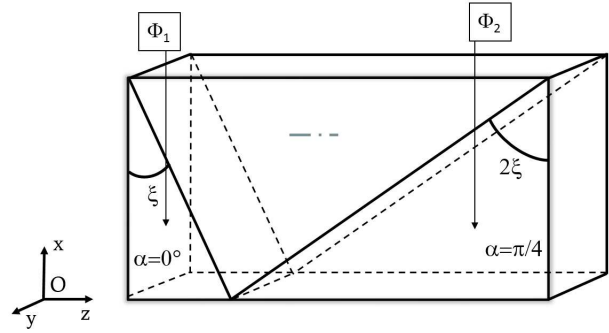}
\caption{Configuration of the wedges to minimize the number of interfaces. The first wedge has an apex angle of $\xi$ and the fast axis of the crystal oriented along the $Oy$ axis. The second wedge has twice that apex angle, $2\xi$, with a fast axis angle of $\frac{\pi}{4}$ with respect to the $Oy$ axis. In between, we insert a piece of the same material (MgF$_2$ in our case) but with its fast axis along the optical axis $Oz$, so that is does not have any birefringent effect. }
\label{onebloc}
\end{figure}

The previous section has shown the need for two birefringent wedges with two different slopes to measure the full Stokes vector. Considering mechanical mounting issues, symmetry around the optical axis and optical aberrations, the two wedges are doubled to create compact rectangular components. The solution proposed here is to suppress one interface for a total equivalent thickness. Figure~\ref{onebloc} shows the configuration of the wedges. The first one, with an apex angle $\xi$ has its fast axis along the horizontal direction $Oy$, while the second one has an apex angle of $2\xi$ and a fast axis angle of $\frac{\pi}{4}$ with respect to the $Oy$ axis. To complete this compact single rectangular component and avoid any refraction or aberration problem, a piece of crystal (the same as used in the 2 wedges, MgF$_2$ in our case) is placed in between. Its fast axis is aligned to the optical axis of the system $Oz$ to suppress birefringent effects.

Using this component, the same calculations as in the previous section lead to an output intensity of:

\begin{multline}
\label{I2}
I_{out}(x,\lambda)=0.5 \cdot [I+Q \cdot \cos{2\Phi}\cos{2\theta} + U \cdot ( \sin{\Phi}\sin{2\Phi}\cos{2\theta} + \\ 
\cos{\Phi}\sin{2\theta} ) +V \cdot \left( \sin{\Phi}\sin{2\theta}-\cos{\Phi}\sin{2\Phi}\cos{2\theta} \right) ]
\end{multline}

with a phase retardance $\Phi$ of: 
\begin{equation}
\label{phi2}
\Phi(x,\lambda)=2\pi\tan{\xi} \cdot \frac{\Delta n(\lambda)}{\lambda} \cdot |x|
\end{equation}

Contrary to the two double-wedge system presented in Sect.~\ref{theory}\ref{global}, here the value of the height $x$ is set to 0 at the top of the component, where light does not cross any birefringent material.

Using this single triple-component offers the same possibilities as the two double-wedge method detailed in Sect.~\ref{theory}\ref{global}, but with the advantage of fewer components and only 4 interfaces, compared to 6. This could be a precious gain in cases when the flux is a crucial issue, e.g., for the observation of cool stars in the FUV. A practical application using this component is presented in Sect.~\ref{arago}.


\section{Application to UVMag/Arago}
\label{arago}

\subsection{Adaptation and design}

In January 2015, the UVMag consortium \cite{uvmag} proposed a UV and visible spectropolarimetry mission, called Arago, to the European Space Agency (ESA). This space project offers a precise polarization measurement over a high-resolution spectrum from 119 to 888~nm \cite{pertenais}. The scientific goal of Arago is to study the entire life of stars and planets, from their formation to their death, as well as feedback into the interstellar medium. Stellar evolution is directly driven by astrophysical processes such as magnetic fields and/or stellar winds. Thanks to a high-resolution spectrograph combined with a precised full Stokes polarization measurement from FUV to near-IR, Arago aims to obtain a full picture of the 3D dynamical environment of stars and planets and their interactions \cite{pertenais14}.

Here, we present an application of the method for spectropolarimetry explained in this article to the Arago instrument, using the single component version of the polarimeter presented in Sect.~\ref{theory}. The system incorporates MgF$_2$ wedges, as this is the only known transparent and birefringent material effective over the full spectral range of Arago, from FUV to NIR (see Fig.~\ref{birefringence}). The apex angle $\xi$ is set to 1.5$^{\circ}$. Equations \ref{I2} and \ref{phi2} are used as the basis for the design of the instrument.
\\

We first determine the spatial period of the different modulation patterns that should be observed in the 2D images and will need to be sampled. A frequency analysis of Eq.~\ref{I2} gives several characteristic periods of the phase retardance $\Phi$. As we have $\cos\Phi$, $\cos{2\Phi}$ and $\cos{3\Phi}$ elements, the possible signals can have $2\pi$, $\pi$ or $2\pi/3$ periods of retardance. We thus need to be able to measure at least one period of the longest signal ($2\pi$) and respect the Nyquist frequency for the shortest signal ($\pi/3$).
\\
\begin{figure}[tbp]
\centering
\includegraphics[width=\linewidth]{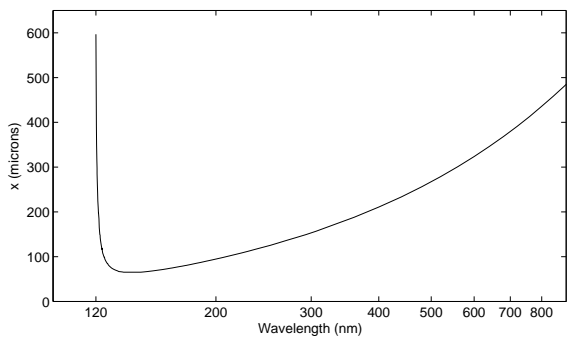}
\caption{Plot of the spatial dimension $x$ over the spectrum, for a given phase retardance. This corresponds to Eq.~\ref{phi2}, with $\xi=1.5^{\circ}$ and $\Phi=\pi/3$. As the birefringence goes to zero around 119.5~nm, the value of $x(\lambda)$ plotted here goes to infinity in the segment [119:120]~nm of the spectrum.}
\label{dx}
\end{figure}

The derivation of Eq.~\ref{phi2} for a phase retardance $\Phi=\pi/3$ is plotted in Fig.~\ref{dx}. The minimum of the curve at $\lambda=145$~nm corresponds to the maximum of the ratio $\Delta n(\lambda)/\lambda$ using the birefringence values presented in Fig.~\ref{birefringence}. At this wavelength, the space gap corresponding to a phase retardance of $\pi/3$ is 65~$\mu$m. This is the minimum value that we have to sample, i.e. the maximum size of the pixels.

To find the maximum value to be sampled, i.e. the height of the wedges, we have to find the wavelength where the ratio $\Delta n(\lambda)/\lambda$ is at a minimum. This occurs at the highest wavelength of study, 888~nm. The part of the spectrum between 119 and 120~nm is not considered as the birefringence goes to zero, as shown in Sect.~\ref{intro}. At $\lambda=888~nm$, the height of the wedges, $x$, has to be large enough to observe at least one period of the lowest frequency signal, $2\pi$. Replacing a $2\pi$ retardance in Eq.~\ref{phi2} at 888~nm gives a minimum height of $x=2.9$~mm.

To summarize, we need to sample the height $x$ of the wedges with at least 65~$\mu$m precision over more than 2.9~mm. In this case we do not undersample the high-frequencies in the UV and we observe at least one period of the low-frequency in the IR part of the spectrum.

The optical design of the high-resolution spectrograph of Arago (spectral resolution of 35000 in the visible and 25000 in the UV spectral range) can be found in Pertenais et al. (2014) \cite{pertenais}. It uses a unique echelle-grating for the entire spectral range, and focuses all of the orders on several detectors. A classical CCD detector with 15~$\mu$m pixel pitch is used to image the 46 orders of the visible spectrum, while two micro-channel plate (MCP) detectors record the 178 short orders in the UV part of the spectrum. The pixel pitch of the MCP detectors is about 20~$\mu$m. As the minimum sample we need to observe is 65~$\mu$m in output of the polarimeter, the spectrometer needs a minimum magnification factor of $1/3.25$. In this way, the precision needed to sample correctly the signal corresponds to the size of a pixel. The consequence of this is that each order will be at least $2.9/3.25=0.892$~mm thick. The value 900~$\mu$m is adopted and is the starting point to define the dimension of the needed detectors.
For the visible detector, 46 orders of 900~$\mu$m with 2 black pixels (15~$\mu$m pixel pitch) between each order lead to a detector of at least 43~mm high (the other direction is defined by the main dispersion of the spectrograph). Visible detectors of this size are common.
The UV part of the spectrum is spread onto two detectors, imaging 89 orders each. The same calculation gives two MCP detectors of at least 84~mm high with 20~$\mu$m pixel pitch. This is a rather large size for MCP detectors.
\\

This section has demonstrated that we can use the proposed method to build a static spectropolarimeter, working over a wide spectral range,  providing all Stokes parameters at once, and adapted to space constraints (single static component). One of the drawbacks of the technique is the need for a large sensitive area in the UV, due to large variation of the ratio $\Delta\lambda(\lambda)/\lambda$ over the spectrum. A possible solution is to modify the echelle grating to decrease the number of orders in the UV, but increase the length of the orders. In this case, the height of the detector needed for the UV domain decreases, and its shape becomes more square.


\subsection{Efficiency}\label{effic}
To compare the efficiency of the Stokes measurement with the concept proposed here to other existing solutions, the formalism developed in del Toro Iniesta et al. (2000) \cite{iniesta} is used.

At a given wavelength (or given column of the image), we consider that the signal is modulated in $n$ states, where $n$ is the number of resolved elements or pixels. For each pixel of the column, we calculate the global Mueller matrix of the polarimeter (see Sect.~\ref{theory}\ref{mueller}). The modulation matrix for this wavelength $O(\lambda)$ is then built with the first lines of the $n$ Mueller matrices. Finally, an optimal demodulation matrix is calculated following:

\begin{equation}
D(\lambda)=\left(O(\lambda)^T \cdot O(\lambda) \right)^{-1} \cdot O(\lambda)^T
\end{equation}

Based on the coefficients of these matrices, the value of the extraction efficiencies of the Stokes parameters can be determined for each wavelength:

\begin{equation}
\epsilon_i=\left(n \cdot \sum_{j=1}^{n} D_{ij}^2 \right)^{-1/2} 
\end{equation}

where $i=[I~~Q~~U~~V]$ and $n$ is the number of pixels used for each wavelength.

The calculations are performed assuming that the modulation is well sampled  over the whole spectrum and that at least one full period is recorded. We find that the efficiencies are almost constant over the spectral range with a small sinusoidal modulation of about 1$\%$. 

\begin{figure}[tbp]
\centering
\includegraphics[width=\linewidth]{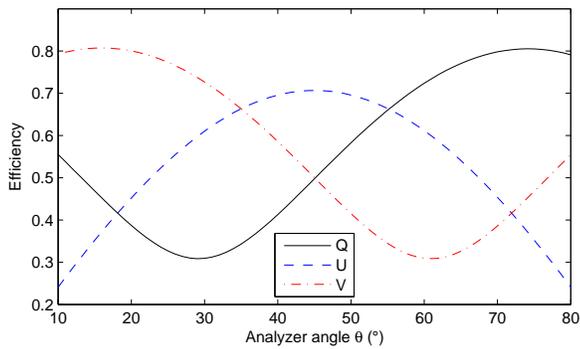}
\caption{Plot of the extraction efficiencies of the Stokes parameters with different polarizer angles $\theta$.}
\label{eff}
\end{figure}

However, the efficiencies depend directly on the angle of the analyzer. Figure~\ref{eff} shows the variation of the efficiencies with the analyzer angle $\theta$. Efficiencies are between 30$\%$ and 80$\%$ depending on the choice of the polarizer angle, using a dual-beam system. Using a simple linear polarizer decreases their values by a factor of 2.

Using the alternative method with rotating plates presented in Sect.~\ref{theory}\ref{global}, we obtain efficiencies varying between 40$\%$ and 60$\%$ over the spectrum, when using a dual-beam system. Therefore, the method presented here and the polychromatric rotating modulator result in comparable efficiencies.

Finally, the choice of the polarizer angle is a free parameter that can be optimised depending on the science goals. For each instrument the angle can be tuned to reach the maximum efficiency for the desired Stokes parameters (see  Fig.~\ref{eff}), e.g. $\theta$=16$^\circ$ if one is interested in circular polarization only (efficiency of 81\% for Stokes V) or  $\theta$=55$^\circ$ to optimize the instrument for linear polarization (efficiency of 66\% for both Stokes Q and U).


\section{Laboratory implementation}
\label{manip}

This section describes the experimental work done to verify the theoretical concept described above. The theoretical and simulated results presented in Sect.~\ref{theory} are compared to actual laboratory observations.

\subsection{Optical design}

To perform the laboratory tests, the first step is to control the entrance light source. To do so, a polychromatic halogen lamp is used in combination with an integrating sphere to ensure the use of unpolarized light. The light leaves the sphere through a pinhole, mimicking a point source, is then collimated and goes through a broadband wire grid linear polarizer to control its polarization state. An optional achromatic quarter waveplate, from the compagny Bernhard Halle Nachfolger GmbH, can be placed to create circular or elliptical polarization states. Finally, a filter wheel offers the possibilities to use monochromatic interferential filters or bandpass filters. The spectral region of interest for the laboratory validation is limited to a part of the visible from about 495 to 725~nm. This spectrum, although wide, is shorter than the one of Arago, for example. However, these tests are performed only to prove the validity and demonstrate a practical application of the proposed concept. As explained in the previous section, the entire instrument can be designed to avoid undersampling at the edges of the spectrum, and permit the measurement of a much wider spectral range.

To build the polarimeter, two double wedges of MgF$_2$ are used as described in Fig.~\ref{wedge}. The wedges, manufactured by the Karl Lambrecht Corporation, have submit angles of 1.5$^{\circ}$ and 3$^{\circ}$, a clear aperture of 10~mm, and fast axis angles of  [$\frac{3\pi}{4}$; $\frac{\pi}{4}$; 0; $\frac{\pi}{2}$]. To complete the polarimetric module of the prototype, a broadband wire grid polarizer is added after the wedges to analyse the polarization state. Its angle is chosen to be $\theta=64^{\circ}$ to (1) avoid particular configurations (see end of Sect.~\ref{arago}\ref{effic}), and (2) obtain 3 different basic 2D patterns for Stokes Q, U and V. Indeed, at some positions of the polarizer, the 3 basic patterns can be mathematically orthogonal for example. By chosing $\theta=64^{\circ}$, we consider the most general case.

\begin{figure}[t]
\centering
\includegraphics[width=\linewidth]{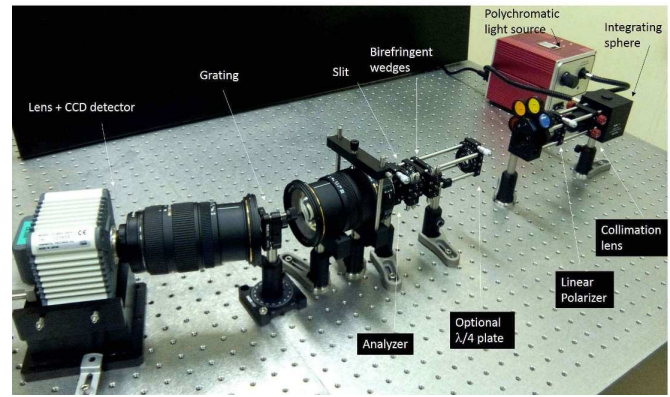}
\caption{Picture of the complete optical bench showing all the elements, from the light source (on the right side) to the detector (on the left side).}
\label{banc}
\end{figure}

After the polarimetric module, a 50~mm focal length photographic lens collimates the light coming from a slit placed right after the birefringent wedges and illuminates a transmissive grating with 600~grooves/mm to disperse the light and create the spectrum. As we do not have any flux problem in the laboratory, the slit is illuminated with a circular collimated beam, even though most of the light is then lost. Another solution would be to place the slit at the focus of a cylindrical mirror or lens. A second 50~mm focal length photographic lens finally focuses the light onto a CCD detector (6.45~$\mu$m pixel pitch and active area of 8.67 x 6.60~mm$^2$). The global view of the optical bench is shown in Fig.~\ref{banc}.

\subsection{Results and comparison to the simulations}

The experimental setup produces a 2D image. This 2D frame is composed of the spectrum in the dispersion direction of the grating (orthogonal to the height of the wedges) and the polarization modulation in the perpendicular direction. By rotating the linear polarizer placed after the light source, the frequencies of the intensity modulation change and we can easily recognize the modulation patterns of the simulations presented in Fig.~\ref{simu}.

In particular, when we input 100\% (Q, U, or V) polarized light, the specific patterns are easily recognized. However, to be able to measure precisely a polarization state, the instrument must be perfectly calibrated. 

\begin{figure}[t]
\centering
\includegraphics[width=\linewidth]{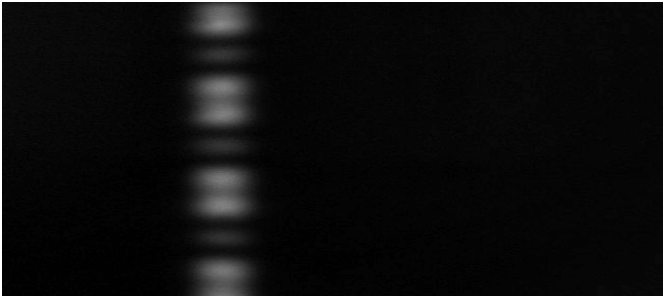}
\caption{Image obtained with a Q entrance polarization state and a narrow-band filter with peak-wavelength at 560~nm and FWHM of 10 $\pm$ 2~nm. This image is used to perform the wavelength calibration of the spectrum.}
\label{calib}
\end{figure}

Interferential filters with different bandwidths and central wavelengths are used to calibrate the position of the spectrum in wavelength on the detector. An example of an image obtained with a narrow-band filter is shown in Fig.~\ref{calib}.

To calibrate the orthogonal direction of the spectrum, the height of the slit (3~mm) is used. In theory, the image on the detector corresponds to a variation of the height $x$ from -1.5 to 1.5 in Eq.~\ref{phi}. For processing reasons (edge effects) and vignetting issues, we start the observation after -1.5~mm and stop before 1.5~mm. We thus cannot simply assume that the 2D frame goes from $x=-1.5$ to $x=1.5$. Instead, the frame obtained in the laboratory is compared (image substraction and cross-correlation) with a theoretical image at given $x_{\rm min}$ and $x_{\rm max}$ values. The $x_{\rm min}$ and $x_{\rm max}$ values corresponding to maximal correlation are extracted and the full frame is calibrated with these values.

\begin{figure}[t]
\centering
\includegraphics[width=\linewidth]{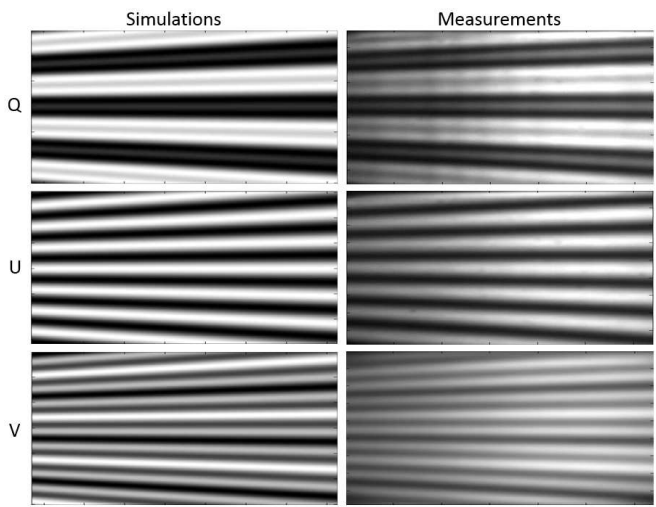}
\caption{Comparison between the theoretical simulations (left panels) described in Sect.~\ref{theory} and the images obtained in practice (right panels) for an entrance polarization of 100\% Q (top), 100\% U (middle) and 100\% V (bottom)..}
\label{comparaison}
\end{figure}

Once the setup is calibrated, the entrance polarization state is modified with the linear polarizer and the achromatic quarter waveplate, and the output images are recorded. The images we obtained are compared with the simulations and presented in Fig.~\ref{comparaison}.

\begin{figure}[t]
\centering
\includegraphics[width=\linewidth]{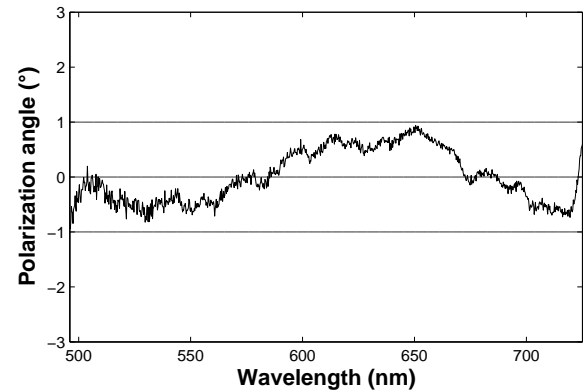}
\caption{The polarization angle measured in the laboratory is plotted for the whole spectral range for a linear polarization of 100\% Q. The measured angle is $0^{\circ}\pm 1^{\circ}$, which corresponds to a measurement error of 1.1\%.}
\label{angle}
\end{figure}

The demodulation code mentioned in Sect.~\ref{theory} is used to measure the Stokes parameters from the 2D images. For a linear entrance polarization we are able to measure the polarization angle with a precision of $\pm 1^{\circ}$ all over the entire spectral range. A plot of the polarization angle value over the spectrum is shown in Fig.~\ref{angle}, for a 100\% Q entrance polarization state (i.e., theorically, a $0^{\circ}$ polarization angle).

\begin{figure}[tbp]
\centering
\includegraphics[width=\linewidth]{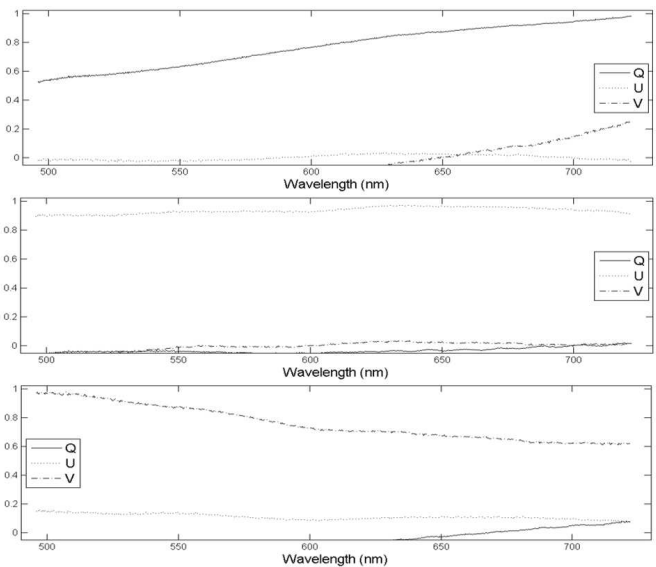}
\caption{Measurements of the normalized Stokes parameters Q/I, U/I and V/I over the spectrum, corresponding to an entrance light with linear polarization state of 100\% Q (top), 100\% U (middle) and 100\% V (bottom).}
\label{stokes_lab}
\end{figure}

In a practical implementation, the final goal of such an instrument is to measure the value of the Stokes parameters over the spectral range. To compensate flat-field and illumination issues, the actual values we determine are Q/I, U/I and V/I for each wavelength of interest. The plots in Fig.~\ref{stokes_lab} show the results measured with the optical bench described above.

The type of entrance polarization state is well retrieved using the demodulation code based on the least-squares method. However, the amount of polarization (100\%) is not properly retrieved.

\subsection{Discussion}

As shown in Fig.~\ref{stokes_lab}, the input Stokes parameters are retrieved with our laboratory setup, but not perfectly. In theory we should measure 100\% Q, U, or V, over the whole spectral range.

First, the value determined is less than 1 (100\%). This is not critical for polarization angle measurements defined by half of the tangent of the ratio between U and Q, but it is important for the determination of the absolute degree of polarization. This is mainly due to the fact that the amplitude of the measured intensity modulation is not maximal. Because of noise, bias, flat-field issues, and performance of the wedges (precise positioning, fabrication imperfections, or non-homogeneity) the "white" stripes of the modulation are not equal 1 and the "black" ones not equal to 0. This loss of contrast propagates in the equations and causes this offset in the absolute Stokes parameter values.

In addition, the determined values of the Stokes parameters are not constant over the spectrum. This is not due to the polarization performances of the components, since they were all chosen to be achromatic. Instead, the demodulation process introduces some errors. The analytic solution to the linear problem (Eq.~\ref{I}) is used to demodulate the signal. However, some uncertainties apply and slightly modify this analytic solution: the detector is not homogeneously illuminated, the birefringence curve depends on the material sample and on temperature, the angle of the polarizer $\theta$ is known with $1^{\circ}$ precision with respect to the entrance polarization state, the apex angle of the wedges $\xi$ and the orientation of the fast axis in each wedge are also not perfectly known.

While the measurements are not perfect, due to the above listed limitations, our experimental setup shows that the concept works and can produce usable results. The current prototype will be difficult to improve because most of the limitations depend on wavelength, time, and temperature, or are inherent to the components. However, a more precise instrument could be built for space missions following this concept.


\section{Conclusions}

The use of this concept offers many advantages that can be applied to various domains. As shown in this article, the idea to spatially modulate the polarization information makes it possible to build a spectropolarimeter without any moving parts and which provides all Stokes parameters simultaneously. Furthermore, the dependence on the wavelength is taken into account, and there is, theoretically, no limitation to the spectral range. The material used for the wedges (through its transparence and birefringence characteristics), in combination with the choice of detectors, will ultimately define the minimum and maximum wavelengths that can be observed. We find that MgF$_2$ is best suited for a wide range from FUV to IR.

The optical bench we set up proves the validity of the theoretical concept. The observational results presented here are comparable to the computed simulations and we retrieve the values of all the Stokes parameters over a given spectral range. The angle of linear polarization is for instance determined with a $\pm 1^{\circ}$ precision.

This concept has thus been demonstrated and could be used in future space missions.
\\

The authors acknowledge support from CNES, R\'egion Midi-Pyr\'en\'ees and Observatoire de Paris.

Martin Pertenais thanks Marion Bonafous for technical support and Laurent Par\`es for supervision, as well as Herv\'e Carfantan and the SISU team in IRAP Toulouse for their help in processing issues. Coralie Neiner and Martin Pertenais also thank William Sparks, Charles Telesco, Frans Snik and Arturo Lopez Ariste for theoretical and technical discussions, and Mary Oksala for language editing.

\end{document}